\documentclass[aps,prl,reprint,superscriptaddress,twocolumn,showkeys,amsmath,amssymb,longbibliography,floatfix]{revtex4-2}
\usepackage{amsmath,amssymb,bbm,mathrsfs,bm,braket,color,graphicx,comment,amsfonts,dsfont}
\usepackage[colorlinks,linkcolor=blue,citecolor=blue,urlcolor=blue]{hyperref}
\usepackage[mathscr]{euscript}
\usepackage{physics}
\usepackage{xcolor}
\usepackage[normalem]{ulem}
\usepackage{bm}
\usepackage{orcidlink}
\usepackage{multirow}
\usepackage{microtype}
\usepackage{bbold}
\usepackage{todonotes}

\input{macros.sty}
\usepackage[capitalise]{cleveref}
\usepackage{times}
\usepackage{printlen}

\newcommand{\1}{\ensuremath{\mathbbm{1}}}

\begin{document}

\title{Topological Gyromorphs}

\author{Laura Gómez Paz\orcidlink{0009-0001-0035-6598}}
\email{laura.gomez-paz@neel.cnrs.fr}
\affiliation{\small Universit\'e Grenoble Alpes, CNRS, Grenoble INP, Institut N\'eel, 38000 Grenoble, France}

\author{Justin Schirmann\orcidlink{0009-0007-7030-0155}}
\email{justin.schirmann@gmail.com}
\affiliation{\small Universit\'e Grenoble Alpes, CNRS, Grenoble INP, Institut N\'eel, 38000 Grenoble, France}

\author{Adam Yanis Chaou\orcidlink{0000-0002-9926-4633}}
\affiliation{Donostia International Physics Center, P. Manuel de Lardizabal 4, 20018 Donostia-San Sebastian, Spain}
\affiliation{Dahlem Center for Complex Quantum Systems and Fachbereich Physik, Freie Universit\"at Berlin, 14195 Berlin, Germany}

\author{Isidora Araya Day\orcidlink{0000-0002-2948-4198}}
\email{isidora@araya.day}
\affiliation{Donostia International Physics Center, P. Manuel de Lardizabal 4, 20018 Donostia-San Sebastian, Spain}
\affiliation{QuTech, Delft University of Technology, Delft 2600 GA, The Netherlands}
\affiliation{Kavli Institute of Nanoscience, Delft University of Technology, 2600 GA Delft, The Netherlands}

\author{Adolfo G. Grushin\orcidlink{0000-0001-7678-7100}}
\email{grushin@dipc.org}
\affiliation{Donostia International Physics Center, P. Manuel de Lardizabal 4, 20018 Donostia-San Sebastian, Spain}
\affiliation{IKERBASQUE, Basque Foundation for Science, Maria Diaz de Haro 3, 48013 Bilbao, Spain}
\affiliation{\small Universit\'e Grenoble Alpes, CNRS, Grenoble INP, Institut N\'eel, 38000 Grenoble, France}

\date{\today}

\begin{abstract}
Gyromorphs are a new class of disordered systems that combine an amorphous-like absence of translational order with quasi-long-range rotational order. Gyromorphs can outperform quasicrystals or hyperuniform arrangements in forming isotropic band gaps, suggesting an avenue to realize robust disordered topological phases.  However, gyromorphs lack exact rotational symmetry, which is only realized on average, posing an obstacle for existing real-space invariants to correctly diagnose topological gyromorphs. In this work we show that gyromorphs can host higher-order topological insulating (HOTI) phases protected by average rotational symmetry, and we develop and systematically compare tools for diagnosing topological phases protected by such symmetry. We introduce symmetry indicators of the effective Hamiltonian based on average rotational symmetries which, when combined with the spectral localizer and a scattering invariant, draw a consistent topological phase diagram.  Our work unlocks gyromorphs as a novel platform to study topological phases beyond crystals, quasicrystals, and amorphous materials.
\end{abstract}

\maketitle

\textit{Introduction} -- 
In the absence of perfect crystalline order, it is still useful to group solids by the action of symmetries on their atomic arrangement.
For example, quasicrystals break translational symmetry while retaining long-range order and rotational symmetries, displaying sharp aperiodic diffraction peaks in their structure factor~\cite{Shechtman:1984kf,Janssen2008}.
In contrast, amorphous solids break translational symmetry retaining short-range atomic order, which smears out diffraction peaks into well-defined rings~\cite{zallen_physics_1998}.

\begin{figure}[h!]
    \centering
    \includegraphics[width=1\columnwidth]{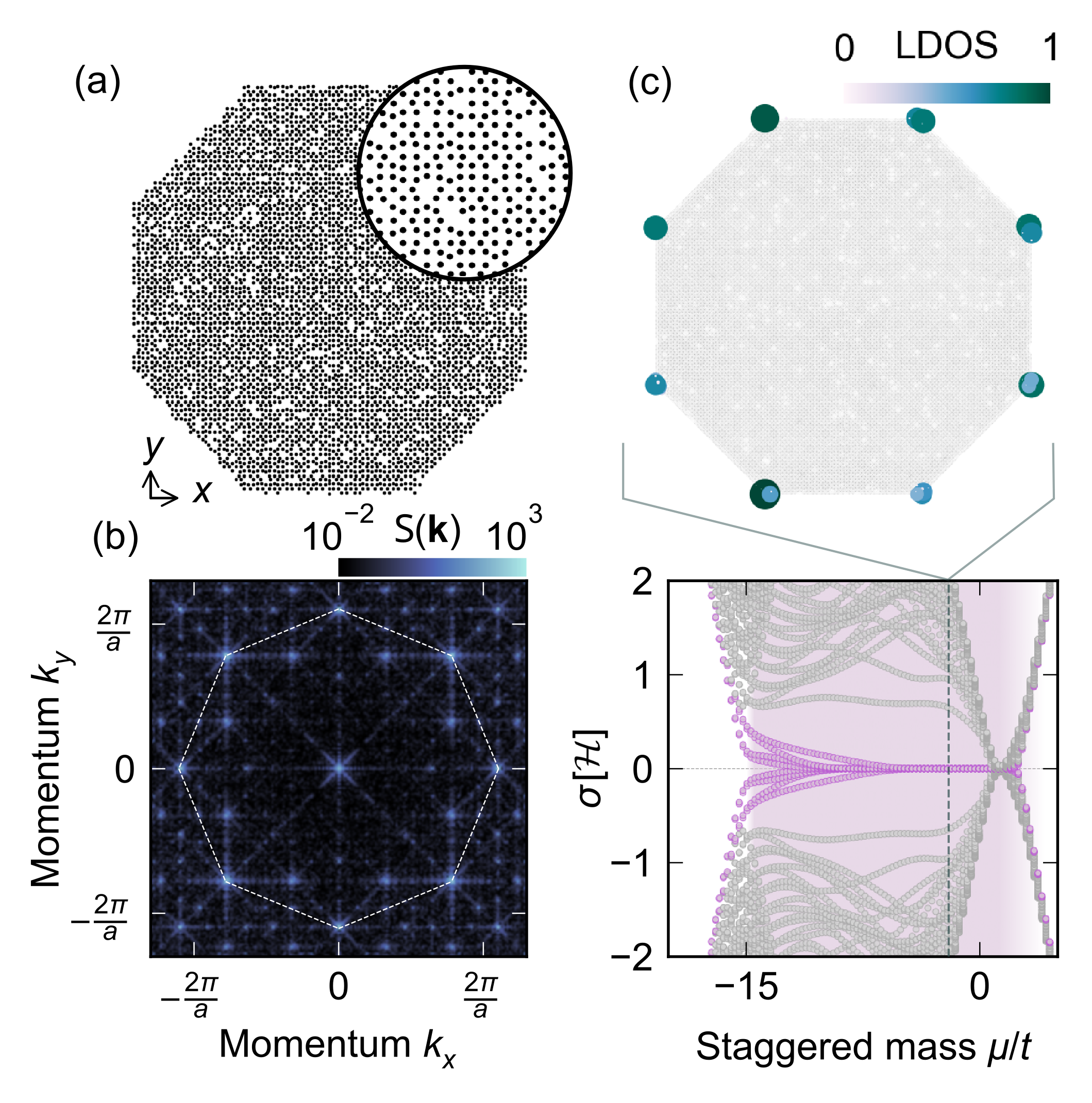}
    \caption{(a) Real-space representation of the $C_8$ locally symmetric gyromorph lattice with $N=5 \times 10^3$ sites.
    (b) Reciprocal-space structure factor of the gyromorph lattice with $N=10^4$. The momentum axes are expressed in units of the average lattice spacing $a= 1/\sqrt{N}$.
    (c) Bottom: 100 smallest eigenvalues of the real-space Hamiltonian \cref{eq:hamgen}, defined on the same lattice as in (b), as a function of the on-site parameter $\mu$. Violet and gray points denote corner and bulk states, respectively. Shading indicates the putative topological region.
    Top: The local density of states (LDOS) of the zero-energy modes is shown for $\mu/t=-2$. }
    \label{fig:Fig1}
\end{figure}

In a recent Letter,  Casiulis \textit{et al.} proposed a new class of disordered materials called Gyromorphs~\cite{Casiulis_gyromorphs_2025}.
Gyromorphs fill the gap between amorphous and quasicrystalline materials by retaining long-range \textit{discrete} rotational order, like quasicrystals, but disposing of long-range radial order, like amorphous systems.
Gyromorphs are the Fourier duals of quasicrystals; the real-space radial distribution function of a gyromorph matches the reciprocal-space features of the structure factor of a quasicrystal, and vice versa~\cite{Casiulis_gyromorphs_2025}. 
Gyromorphs display average local and average global rotational symmetries, as illustrated in \cref{fig:Fig1} with a $C_8$-symmetric gyromorph (a) and its sharp diffraction peaks (b).

The average rotational symmetries inherent to gyromorphs force us to revisit how we diagnose non-crystalline topological phases.
Existing real-space topological methods~\cite{Kitaev20062,ceresoli_orbital_2007,bianco11,Loring2010,Prodan2010,Prodan2011,marrazzo_locality_2017,LORING2015,Akagi2017,Loring2019,marsal_topological_2020,corbae_evidence_2020,Markov2021,Schulz-Baldes21,Hannukainen2022,Ornellas2022,spillage_2022,jezequel2023modeshell,favata_single-point_2023,Schulz-Baldes2024,Franca2024b,Hannukainen2024,uria-alvarez_deep_2022,Cerjan_local_2024,Cerjan_tutorial_2024,Li2024,Bau2024,Bau2024b,jezequel2025modeshell,Chaou2025,Zijderveld_2025, Bianco_Orbital_2013, Seleznev_towards_2023, Chen2023, Oliveira_Robustness_2024}
diagnose topological phases in amorphous~\cite{agarwala_topological_2017,mansha_robust_2017,xiao_photonic_2017,Mitchell2018,poyhonen_amorphous_2018,Prodan2019,corbae_evidence_2020,marsal_topological_2020,Grushin2020,Agarwala_HOTI_2020,focassio_structural_2021,wang_structural-disorder-induced_2021,Spring_amorphous_2021,Manna2022,Junyan2022,marsal_obstructed_2022,manna_noncrystalline_2022,Wulles2022,spring_isotropic_2026,Grushin2023,Cassella2023,Corbae_2023,Cheng23,Franca2024,Bera2024,Martinez2026,uria2025}
and quasicrystalline~\cite{Kraus2012,Tran:2015cj,Bandres:2016gx,Fuchs:2016hp,Huang2018,Huang2018b,Fuchs:2018dd,Loring2019,Varjas_2019,Chen2019,He2019,Duncan2020,Fan2021,Zilberberg:21,Hua2021,Else2021,Jeon2022,JustinHat,manna_noncrystalline_2022,Roche2025,caiger2026fractaltopologymajoranabound} systems
when they are protected by exact on-site (e.g., time-reversal, particle-hole, and chiral symmetries) and spatial symmetries. 
However, for disordered topological systems protected by average spatial symmetries, computing a real-space invariant that accounts for the
symmetry remains challenging. Existing approaches require either imposing global spatial symmetry of a single realization~\cite{Zijderveld_2025,zijderveld_2026,Tao_average_2023}, or restoring the symmetry via an ensemble average over  realizations~ \cite{Spring_amorphous_2021,Tao_average_2023,schirmann_geometry_2025},  
or considering models with exact local spatial symmetry~\cite{corbae_evidence_2020,marsal_topological_2020,marsal_obstructed_2022,manna_noncrystalline_2022}.
This methodological limitation leaves topological gyromorphs uncharacterized, making it difficult to identify the underlying topological phases in experimental realizations.

Gyromorphs are able to outperform quasicrystals and hyperuniform systems in forming clean isotropic spectral gaps~\cite{Casiulis_gyromorphs_2025}, an advantageous property for photonic applications~\cite{Vynck2023,Milosevic2019,Man2013,Piechulla2021}, and to enhance topological robustness compared to other disordered media.
Moreover, naturally grown amorphous materials are in general isotropic, realizing $C_\infty$ symmetry on average~\cite{Spring_amorphous_2021, spring_isotropic_2026,schirmann_geometry_2025}, and hence are not ideal candidates to host non-crystalline higher-order topological phases~\cite{Agarwala_HOTI_2020,Tao_average_2023}.
In contrast, gyromorphs have average discrete rotational symmetry $C_n$ built in locally in a single realization, and show well developed spectral gaps without trivially localized in-gap states, features we exploit in this work.

In this work, we demonstrate that gyromorphs host higher-order topological phases and solve the problem of how to identify them. We consider a Hamiltonian on the gyromorph shown in \cref{fig:Fig1} with local $C_8$, particle-hole and time-reversal symmetries and demonstrate that it hosts a higher-order topological phase with eight corner modes.
We introduce symmetry indicators of the effective Hamiltonian based on average rotational symmetries. By combining them with spectral localizer and a scattering invariant~\cite{Fulga_2012, Zijderveld_2025}, we establish a consistent topological phase diagram.

\textit{Gyromorph tight-binding model} -- 
To construct the gyromorph we generate a point set with average local rotation symmetry using the Fast Reciprocal-Space Correlator (FReSCo)~\cite{Shih_fast_2024}, an optimization-based method that enforces prescribed features in the point-set’s structure factor.
We run the optimization with a set of $N$ uniformly distributed points in two dimensions and impose $n$ rotationally symmetric Bragg peaks in the structure factor at a fixed radial distance $2\pi/a$ from the $\Gamma$ point, with $a = 1/\sqrt{N}$.
We then select an octagonal region from the resulting point set, as depicted in \cref{fig:Fig1}(a). Its structure factor is shown in \cref{fig:Fig1}(b).

On the gyromorph point-set we use the same real space Hamiltonian as in Ref.~\cite{Varjas_2019}, a pair of oppositely spin-polarized topological superconductors, each in class D,
\begin{equation}
\label{eq:hamgen}
    \mathcal{H} = \sum_{j} \bm{c^{\dagger}_j} \epsilon_j \bm{c_j} + \sum_ {j,k}  f(r_{jk}^{\phantom{}}) \bm{c^{\dagger}_j} T_{jk}^{\phantom{}}  \bm{c_k^{\phantom{}}}
    ,
\end{equation}
with $\bm{c^{\dagger}_j}=(c^{\dagger}_{j\uparrow},c_{j\uparrow},c^{\dagger}_{j\downarrow},c_{j\downarrow})$, $c^{\dagger}_{j,\sigma}$ the creation operator at site $j$ with spin $\uparrow,\downarrow$,  $\epsilon_j$ the on-site potential, $T_{jk}$ the all-to-all hopping matrix, and $f(r_{jk})$ the hopping strengths as a function of the inter-atomic distances $r_{jk}=|\mathbf{r}_j-\mathbf{r}_k|$.
The Pauli matrices $\sigma$ and $\tau$ act on the spin and electron-hole degrees of freedom, respectively.
To ensure that the model respects the average local $C_n$ symmetry of the gyromorph point-set we use the model in Ref.~\cite{Varjas_2019} 
for a second-order two-dimensional topological insulator~\cite{Benalcazar2017ScienceMultipole,Benalcazar2017PRBMultipoleMoments,Schindler2018SciAdvHOTI,Trifunovic2019PRXHOBBC,Langbehn2017PRLReflectionSecondOrder,Geier2018PRBOrderTwoCrystalline,Franca2018PRBAnomalousHOTI,Xie2021}
\begin{eqnarray}
\nonumber
    T_{jk} &=&  \frac{t}{2} \sigma_z\otimes\left( \tau_z - i \left[\tau_x \cos(\theta_{jk}) + \tau_y \sin(\theta_{jk})\right]\right) \\
    \label{eq:RealSpaceHoppings}
    &+&\frac{V}{2}\cos(\dfrac{n}{2} \theta_{jk}) \sigma_y\otimes\tau_0 ,\\
    \label{eq:onsite}
    \epsilon_j &=& \mu\left(\sigma_z \otimes \tau_z\right),
\end{eqnarray}
where $t$ is the hopping strength
~\footnote{Compared to Ref.~ \cite{Varjas_2019} we directly choose the superconducting pairing, $\Delta$ in Ref.~\cite{Varjas_2019}, to be equal to the hopping $t$.}
, $\mu$ is a staggered mass potential, $\theta_{jk}$ corresponds to the bond angle between the site $j$ and the site $k$, and $V$ couples opposite spins. 
Additionally, to ensure a local Hamiltonian we choose $f(r_{jk})$ to decay exponentially with the separation between sites $r_{jk}$~\cite{agarwala_topological_2017} 
\begin{equation}
    f\left(r_{jk}\right) = \exp\left(1-r_{jk}/R\right)\Theta(R-r_{jk}),
    \label{eq:f-function}
\end{equation}
where $R$ is the cutoff for the hoppings with a value proportional to $1/\sqrt{N}$ and $\Theta$ is the Heaviside function.
The Hamiltonian in Ref.~\cite{Varjas_2019} is composed of two spin-polarized layers and preserves $C_n M$
\begin{subequations}
\label{eq:cnm}
\begin{eqnarray}
\label{eq:c8}
C_n &=& \exp(-i \frac{\pi}{n} \sigma_0\otimes \tau_z), \\
\label{eq:M}
M &=& \sigma_z \otimes \tau_0,
\end{eqnarray}
\end{subequations}
where $C_n$ is an n-fold rotation within a layer and $M$ is a mirror operator that maps one layer to the other.
The last term in \eqref{eq:RealSpaceHoppings} forces $n$ to be a multiple of four; we fix $n=8$ in what follows. 
The gyromorphic Hamiltonian, therefore satisfies average $C_8M$ symmetry and exact particle-hole ($\mathcal{P}^2=+1$), time-reversal ($\mathcal{T}^2=+1$) and chiral symmetries, placing it in class BDI. 
The symmetry's representations are detailed in Appendix~\ref{app:symmetries}.

Figure~\ref{fig:Fig1}(c) shows the energy spectrum for the gyromorph Hamiltonian as a function of $\mu$. For values of parameters $t=1$, $V=2$, $R=2.265a$, and $N = 10^4$, we observe a range of $\mu$ for which there are eight in-gap states localized at the corners of the octagonal sample, as shown in the top panel. 
Away from two gap closings transitions at $\mu\sim -15$ and $\mu\sim 2$, the spectrum (excluding corner states) retains a clean gap.
This contrasts findings in similar models of amorphous systems, where the spectral gaps close because of disorder~\cite{Agarwala_HOTI_2020,Tao_average_2023}. This advantageous feature, seen in photonic gyromorphs~\cite{Casiulis_gyromorphs_2025}, results in well-defined bulk-gaps in the electronic spectrum, seen in \cref{fig:Fig1}.

\textit{Average symmetry indicators} -- 
Our next goal is to establish if the in-gap corner modes are topologically protected. 
To do so we develop a symmetry indicator that explicitly uses the average $C_8$ symmetry of the gyromorph.
Our approach is based on a momentum-space effective Hamiltonian~\cite{Varjas_2019} which has already shown promise to diagnose topology in amorphous systems~\cite{marsal_topological_2020,marsal_obstructed_2022,Casiulis_gyromorphs_2025}. It is defined as
\begin{align}
H_{\mathrm{eff}}(\bm k) &= \frac{1}{2}(G_{\mathrm{eff}}^{-1}(\bm k) + [G_{\mathrm{eff}}^{-1}(\bm k) ]^{\dagger} ).
\label{eq:effectiveH}
\end{align}
where $ G_{\mathrm{eff}}(\bm k)_{l,l'} = \langle \bm k, l \lvert G \rvert \bm k, l' \rangle$ is the zero-energy Green's function $G = \lim_{\eta \to 0} (\mathcal{H} + i\eta)^{-1}$ projected onto normalized plane-waves $\lvert \bm k, l\rangle$ for each momentum $\bm k$ and orbital $l$.
Based on the effective Hamiltonian~\cite{Varjas_2019}, Refs.~\cite{corbae_evidence_2020,marsal_topological_2020,marsal_obstructed_2022} developed symmetry indicators for spatial symmetries (including rotations) that are locally exact.

One advantage of the effective Hamiltonian is that it has a `disorder averaging' effect in the thermodynamic limit, restoring average (local) symmetries of the system to exact symmetries of $H_\textrm{eff}$. This enables the identification of topological phases using symmetry eigenvalues of $C_8M$ at high-symmetry momenta. The absence of translation symmetry in gyromorphs poses a new problem: at which momenta should one look for band inversions? The lack of a reciprocal lattice prevents us from identifying high-symmetry points at the edge of the Brillouin zone.
We observe that we may still use the $C_8M$ operator to find the momenta that minimize the norm of $[H_{\mathrm{eff}}(\mathbf{k}), C_8 M]$ thus revealing the effective high-symmetry-like momenta.

\begin{figure}
    \centering
    \includegraphics[width=1\columnwidth]{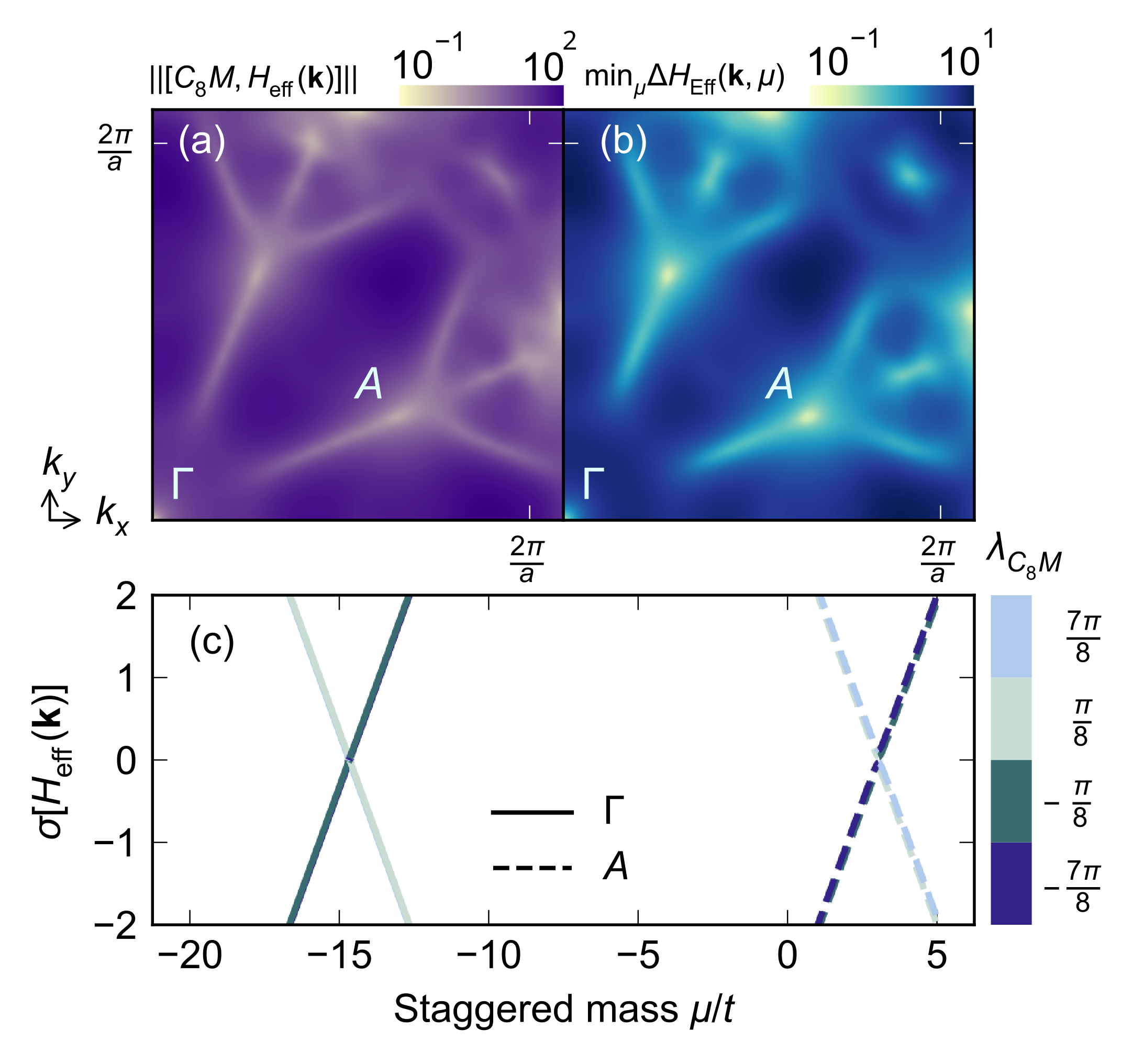}
    \caption{(a) Norm of the commutator $[H_{\mathrm{eff}}(\mathbf{k}), C_8 M]$ from \cref{eq:effectiveH,eq:cnm} in momentum space. (b) Minimum bulk gap of the effective Hamiltonian in momentum space for $\mu/t \in [-20, 5]$, in logarithmic scale. The points $\Gamma$ and $A$ correspond to the high-symmetry momenta where the bulk gap closes. (c) Spectrum of the effective Hamiltonian computed at the $\Gamma$ and $A$ points represented by solid and dashed lines, respectively, as a function of the mass parameter $\mu / t$. The color encodes the $C_8M$ eigenvalues associated with each subspace. We see two crossings as a function of $\mu/t$, each involving a different pair of symmetry eigenvalues.}
    \label{fig:Fig2}
\end{figure}

\cref{fig:Fig2}(a) shows the norm of said commutator over a range of momenta, and \cref{fig:Fig2}(b) shows the minimal gap $\Delta$ of $H_{\mathrm{eff}}(\mathbf{k})$ for all values of $\mu$.

Using the hopping cutoff $R=2.265a$, we observe that the effective Hamiltonian's gap only closes at the high-symmetry-like momenta $\Gamma=(0,0)$ and $A=(1.3\pi/a,0.53\pi/a)$, which are stable as a function of $\mu/t$.
Points related to $A$ by $C_8$ carry no additional information.
Gap closings also occur at high-symmetry momenta with $|\mathbf{k}| > 2\pi/a$, but as the disorder-induced self energy becomes increasingly important at momenta beyond $|\mathbf{k}| \sim 2\pi/a$, predictions based off $H_{\mathrm{eff}}(\mathbf{k})$ become unreliable~\cite{schirmann_geometry_2025}. 
Hence we do not consider them in the symmetry analysis.

Based on these considerations we construct the average symmetry indicators focusing on $H_{\mathrm{eff}}(\Gamma)$ and $H_{\mathrm{eff}}(A)$.
At these points, validated by \cref{fig:Fig2}(a), we can label the Hamiltonian bands by $C_8M$ eigenvalues for different values of $\mu/t$, as shown in \cref{fig:Fig2}(c).
The bulk-gap closings for these momenta \cref{fig:Fig2}(b) happen at values of $\mu/t$ close to where the spectrum of the real-space Hamiltonian in \cref{fig:Fig1}(c) is gapless.
We observe symmetry eigenvalue inversions as a function of $\mu/t$, at $\Gamma$ for $\mu/t\sim-15$ and at $A$ for $\mu/t\sim 2$; signaling topological phase transitions at both gap closing points~\cite{Fu2007,Kruthoff2017,po2017symmetry,Khalaf2018}. 

\textit{Real-space invariants} -- 
The inversion of symmetry eigenvalues of $H_{\mathrm{eff}}$ confirms that the phase within $-15<\mu/t<2$ and outside this interval are topologically distinct phases of matter. 
To unequivocally demonstrate a higher-order topological phase what remains is to apply real-space invariants that probe local symmetries, because higher-order topological insulators in $2d$ require either particle-hole or chiral symmetries to protect corner modes.
We demonstrate that the gyromorph may realize either higher-order topological phases by applying existing but complementary approaches: a symmetry-reduced spectral localizer~\cite{Cerjan_local_2024,Cerjan_tutorial_2024} and a scattering invariant~\cite{Zijderveld_2025}.
We choose the former to diagnose a chiral-symmetry protected phase, and the latter to diagnose a phase protected by particle-hole and eight-fold rotation symmetry~\footnote{We note that we could have used these two methods with either of the particle-hole or chiral symmetries}.

The spectral localizer~\cite{Loring2010,Cerjan_tutorial_2024} 
is a real-space operator whose spectral properties enable the formulation of topological invariants by quantifying how far the position and Hamiltonian operators are from commuting.
Besides discrete symmetries, the localizer formalism allows to treat crystalline symmetries like mirror or inversion symmetries~\cite{Cerjan_local_2024}, although it does not currently incorporate rotational symmetries. In its current form the localizer only serves as a probe for \emph{extrinsic} higher-order phases~\cite{Sitte2012,Geier2018PRBOrderTwoCrystalline}. Topological transitions in extrinsic HOTIs require either a surface or bulk gap closing and therefore do not require a topological bulk. \emph{Intrinsic} HOTIs, on the other hand, require a bulk gap closing to undergo a topological phase transition and only arise in systems with a topological bulk. This further motivates the necessity of the symmetry-indicator approach developed in \cref{fig:Fig2}.


Nevertheless, we may still use the localizer formalism to diagnose topology stemming from discrete symmetries of \cref{eq:hamgen},
in our case an anticommuting chiral operator $\mathcal{S} = \1_N \otimes \sigma_x\otimes\tau_0$.
To this end we use the symmetry-reduced spectral localizer~\cite{Cerjan_local_2024}
\begin{align}
    \label{eq:symredloc}
    \tilde{L}_{\mathbf{x}} = [i \mathcal{H} + \kappa\,\hat{e}_d\cdot(\mathbf{X} - \mathbf{x}\1) ]\mathcal{S},
\end{align}
which chooses a combination of position operators $\mathbf{X} =(X,Y)$ along the direction of the unit vector $\hat{e}_d$, $\kappa$ as a tuning parameter that ensures consistent units, $\1$ as shorthand notation for $\1_N \otimes \sigma_0\otimes \tau_0 $, and $\mathbf{x}$ as the position to compute the localizer.

We can use Eq.~\eqref{eq:symredloc} to signal topological phase transitions through its half-signature
\begin{align}
    \label{eq:idxlocalizer}
    \zeta_{\mathbf{x}} = \frac{1}{2}\mathrm{sig}(\tilde{L}_{\mathbf{x}}),
\end{align}
defined as the half-difference between positive and negative eigenvalues.
To demonstrate how, we ask what is the action of $\tilde{L}_{\mathbf{x}}$ on a spatially localized zero-mode of $\mathcal{H}$, such as potential corner-modes.
By definition, the first term in Eq.~\eqref{eq:symredloc} acts trivially on a zero-mode of $\mathcal{H}$. 
The second term, for small $\kappa$, shifts the zero mode slightly up or down from zero, depending on whether the zero mode has positive or negative chirality, and whether it is positioned to the left or right of the chosen $\mathbf{x}$ along $\hat{e}_d$.
Hence, as we tune $\mathbf{x}$ we will match the projected position of a corner state along $\hat{e}_d$, and a localizer eigenvalue will cross zero, changing the localizer's half-signature.

\begin{figure}
    \centering
    \includegraphics[width=\columnwidth]{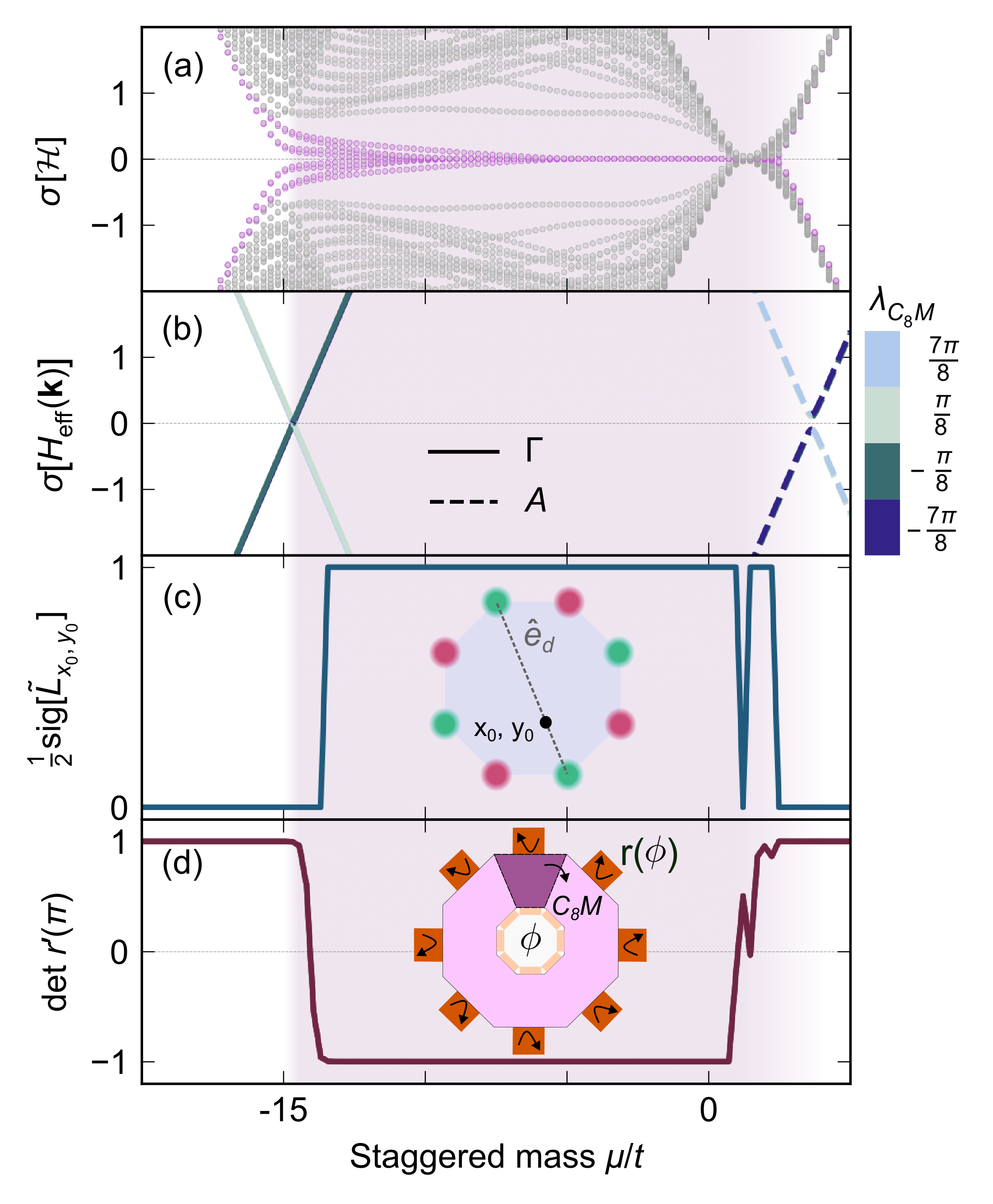}
    \caption{
    Topological phase diagram as a function of the staggered mass $\mu/t$. (a) Real-space Hamiltonian spectrum from \cref{fig:Fig1}(c). (b) Inversion of the $C_8 M$ eigenvalues at the gap closings of the effective Hamiltonian at the high-symmetry momenta from \cref{fig:Fig2} (c) Half-signature of symmetry reduced localizer from \cref{eq:idxlocalizer} for
    $\kappa=0.5$, and $\mathbf{x}=(x_0, y_0)$ shown in the inset. Green and red circles denote the positive and negative chiral charges of the corner modes. The numerical instability at $\mu/t\sim 0$ is consistent with a small gap of both the Hamiltonian and spectral localizer, see Appendix \ref{App:reducedlocalizer}. (d) Higher-order scattering invariant from \cref{eq:scattering_invariant}, computed using the eight-fold symmetric transport setup shown in the inset. The outer leads are shown in dark orange while inner leads are colored by light orange. The dark pink slice of the octagon is mapped by $C_8M$ through all the sample, shown in pink.
    }
    \label{fig:Fig3}
\end{figure}

This implies that the half-signature counts the net chiral charge of zero-energy modes projected onto the direction along $\hat{e}_d$ at the specific location $\mathbf{x}$ (see \cref{App:reducedlocalizer} and Refs.~\cite{Cerjan_local_2024,Cerjan_tutorial_2024}). 
By choosing $\hat{e}_d$ along a direction connecting two opposite vertices of the octagon and $\mathbf{x} = (x_0, y_0)$ on that line, as depicted in the inset of \cref{fig:Fig3}(c), we can then signal a topological transition. 
This expectation is confirmed by \cref{fig:Fig3}(c) that shows Eq.~\eqref{eq:idxlocalizer} changing at the values of $\mu/t$ matching those where there is a Hamiltonian bulk-gap closing, shown in \cref{fig:Fig3}(a), and when the $C_8$ symmetry eigenvalues invert, shown in \cref{fig:Fig3}(b).

As an additional confirmation of the topological phase diagram of the gyromorph, we use the scattering theory of higher-order topological phases~\cite{Zijderveld_2025}.
In contrast to the localizer, this has the added advantage that it directly probes bulk topology and thus identifies intrinsic higher-order phases rather than just extrinsic ones.
Differently from the effective Hamiltonian, the scattering invariant does not probe bulk-gap closings, but rather delocalization transitions of the bulk states at the Fermi level by using a transport setup~\cite{Fulga_2012}.
This probe, however, requires the exact symmetries of the Hamiltonian to hold in order to constrain the scattering equations, so we enforce an additional global $C_8 M$ symmetry of the system.
Nevertheless, introducing such exact global symmetry retains 
the average local symmetries that protect the bulk topology
~\cite{zijderveld_2026}.

We thus construct the scattering setup by considering a $1/8$-slice of the gyromorph model and applying $C_8M$. 
We remove all sites inside a radius from the center of the sample, and attach a pair of $C_8M$-symmetric leads to the inner and outer radius of the sample.
Then we thread a flux $\phi$ through the hole and compute the reflection matrix $r(\phi)$ from the inner lead, as shown in the inset of \cref{fig:Fig3}(d).

The topological invariant follows from the procedure developed in Ref.~\cite{Zijderveld_2025} and we briefly summarize it here.
We start by constraining the reflection matrix using the particle-hole $\mathcal{P}$ and $C_8M$ symmetries of the Hamiltonian:
\begin{align}
    C_8M: \quad r(\phi) &= Q_{C_8M} r(\phi) V^{\dagger}_{C_8M} \\
    \mathcal{P}: \quad  r(\phi) &= r^\ast(-\phi),
\end{align}
where $Q_{C_8M}$ and $V_{C_8M}$ are matrices that act on the incoming and outgoing modes, respectively.
The first constraint allows us to block-diagonalize the reflection matrix into eight blocks, while the second constraint allows us to make $r(\phi)$ real at $\phi = 0, \pi$.
At $\phi = \pi$, the $\pm 1$ subspaces of the $C_8M$ operator commute with particle-hole symmetry, allowing us to define
\begin{equation}
    \nu = \det r'(\pi),
    \label{eq:scattering_invariant}
\end{equation}
where $r'$ is a real block of the reflection matrix corresponding to the $+1$ eigenvalue of $C_8M$.
Additionally, because a gapped sample must have a unitary reflection matrix, Eq.~\eqref{eq:scattering_invariant} may only take values $\pm 1$: $\nu = -1$ signals a topological phase, while $\nu = 1$ corresponds to a trivial phase.
Similarly to a Hamiltonian invariant, the only way to change the value of $\nu$ is by undergoing a delocalization transition of the bulk states at the Fermi level or by breaking the symmetries that constrain the scattering equations. \cref{fig:Fig3}(d) shows the result as a function of $\mu/t$, whose transitions agree with the rest of \cref{fig:Fig3}.


\textit{Conclusions} -- Gyromorph's ability to retain well-defined gaps with topological corner modes while being intrinsically disordered suggests that they are a promising platform to realize robust non-crystalline higher-order topological phases. 
We were able to demonstrate a higher-order topological phase in a gyromorph despite its lack of global rotational symmetry. To do so, we have developed symmetry indicators based only on average local rotational symmetry, labeling the topological phase transition with average $C_8$ eigenvalues. Then we used the spectral localizer and scattering invariants to confirm the higher-order topological phase, as this phase requires an additional chiral or particle-hole symmetry. Taken together our combined methodology correctly diagnoses higher-order topological phases protected by gyromorphic average local rotational symmetry. Although the Hamiltonian construction explored here is limited to $C_n$ symmetries where $n$ is a multiple of four, the methodological tools we developed for identifying topological phases—the average symmetry indicators, spectral localizer, and scattering invariant—apply to any $C_n$ symmetry that can protect a higher-order topological phase.
This makes our framework broadly applicable. In particular, this methodology is applicable to quasicrystals, such as the one studied in Ref.~\cite{Varjas_2019}. With this development gyromorphs now exist as a novel class of disordered materials that can host topological phases. 

Topological photonic systems \cite{lutopological2014,Bandres:2016gx} are a natural platform to explore and take advantage of topological gyromorphs, as this is the context where they were proposed~\cite{Casiulis_gyromorphs_2025}, and higher-order topology has been experimentally realized~\cite{schulz2022photonic}. Other platforms, like CO molecules on metallic surfaces, have already realized higher order phases without crystalline symmetry, and might be appropriate to realize topological gyromorphs~\cite{Kempkes2019}. We have restricted ourselves to two-dimensional gyromorphs but they exist in three-dimensions~\cite{Casiulis_gyromorphs_2025}, as do photonic realizations of higher-order topological insulators~\cite{Wang2025}. In conjunction with our work, these may allow the design of electromagnetic wave propagation in disordered media with well defined band-gaps~\cite{Vynck2023}. Exploring topological gyromorphs along these directions is an interesting avenue for future work.

\textit{Acknowledgments} --
We thank A. Cerjan and C. Fulga for insightful discussions about the spectral localizer, and A.~R.~Akhmerov and D.~Varjas for help on algorithmic optimizations. We also thank A.~R.~Akhmerov for useful discussions about the scattering invariant.
The authors acknowledge the technical and human support provided by the DIPC Supercomputing Center. Computational resources were provided by the Hyperion cluster at the Donostia International Physics Center (DIPC).
All authors are supported by the European Research Council (ERC) Consolidator grant under grant agreement No. 101042707 (TOPOMORPH). 
A.~Y.~C. acknowledges financial support from the Deutsche Forschungsgemeinschaft (DFG, German Research Foundation), project number 277101999, CRC TR 183 (projects A03).
I.~A.~D. acknowledges financial support from the Netherlands Organization for Scientific Research (NWO/OCW) as part of the Frontiers of Nanoscience program.

\textit{Code availability} -- The codes and data to reproduce the figures in this work can be found in Ref.~\cite{gyro_zenodo2026}. Numerical calculations were performed using the Kwant code~\cite{Groth_2014_kwant}.

\bibliography{gyro.bib}

\clearpage
\newpage

\setcounter{secnumdepth}{5}
\renewcommand{\theparagraph}{\bf \thesubsubsection.\arabic{paragraph}}

\renewcommand{\thefigure}{S\arabic{figure}}
\setcounter{figure}{0}

\appendix

\section*{End Matter}

\section{Symmetries of the model}
\label{app:symmetries}
The model introduced in the main text possesses time-reversal, particle–hole, and chiral symmetries. In addition, the effective Hamiltonian exhibits an average $n-$fold rotational symmetry, inherited from the structure of the underlying point set. In this Appendix, we present the explicit form of the corresponding symmetry operators for $n=8$ and detail the transformation properties of both the real-space Hamiltonian and the effective Hamiltonian under these symmetries.
\subsection{Particle-hole symmetry}
The system's particle-hole symmetry operator $\mathcal{P}$ is 
\begin{equation}
\label{eq:ParticleHole}
\mathcal{P} =  \1_N \otimes \sigma_0\otimes\tau_x,
\end{equation} 
where $\1_N$ denotes the identity matrix acting in position-space. Under particle-hole symmetry, the real-space Hamiltonian transforms as
\begin{equation}
    \mathcal{H} = -\mathcal{P}\mathcal{H}^{*}\mathcal{P}^{-1},
\end{equation}
while the effective Hamiltonian transforms as
\begin{equation}
    H_{\mathrm{eff}}(\mathbf{k}) = -U_\mathcal{P}H_{\mathrm{eff}}(-\mathbf{k})^{*}U_\mathcal{P}^{-1},
\end{equation}
where
\begin{equation}
    U_\mathcal{P} = \sigma_x\otimes\tau_x.
\end{equation}
\subsection{Time-reversal symmetry}
The model is also invariant under time-reversal symmetry. The corresponding operator $\mathcal{T}$ is defined as:
\begin{equation}
\label{eq:TimeReversal}
    \mathcal{T} =  \1_N \otimes\sigma_x\otimes\tau_x,
\end{equation}
where $\1_N$ denotes the identity matrix acting in position-space. Under this symmetry, the real-space Hamiltonian satisfies
\begin{equation}
    \mathcal{H} = \mathcal{T}\mathcal{H}^{*}\mathcal{T}^{-1},
\end{equation}
and the effective Hamiltonian transforms as
\begin{equation}
    H_{\mathrm{eff}}(\mathbf{k}) = U_\mathcal{T}H_{\mathrm{eff}}(-\mathbf{k})^{*}U_\mathcal{T}^{-1},
\end{equation}
where
\begin{equation}
    U_\mathcal{T} = \sigma_x\otimes\tau_x.
\end{equation}
\subsection{Chiral symmetry}
By combining the particle--hole symmetry operator~\cref{eq:ParticleHole} and the time-reversal symmetry operator~\cref{eq:TimeReversal}, one can define a chiral symmetry operator given by
\begin{equation}
    \mathcal{S} = \1_N \otimes \sigma_x\otimes\tau_0,
    \label{chiralsym}
\end{equation}
where $\1_N$ denotes the identity matrix acting in position-space. The Hamiltonian then satisfies
\begin{equation}
    \mathcal{H} = - \mathcal{S}\mathcal{H}\mathcal{S}^{-1}.
\end{equation}
The effective Hamiltonian transforms as
\begin{equation}
    H_{\mathrm{eff}}(\mathbf{k}) = -U_\mathcal{S}H_{\mathrm{eff}}(\mathbf{k})U_\mathcal{S}^{-1},
\end{equation}
where
\begin{equation}
    U_\mathcal{S} = \sigma_x\otimes\tau_0.
\end{equation}

\subsection{$C_8M$ symmetry}
Although the gyromorph possesses neither an exact global eightfold ($C_8$) rotational symmetry nor a mirror ($M$) symmetry, the effective Hamiltonian nonetheless inherits an approximate $C_8M$ symmetry arising from the local eightfold symmetry of the underlying point configuration. The corresponding symmetry operator is given by
\begin{equation}
    C_8M = \exp\left(-i\frac{\pi}{8}\sigma_0\otimes\tau_z\right) \sigma_z \otimes \tau_0.
\end{equation}
Under this symmetry, the effective Hamiltonian transforms as
\begin{equation}
    H_{\mathrm{eff}}(\mathbf{k})
    = \left(C_8M\right)\,
      H_{\mathrm{eff}}\!\left(R(\pi/4)\cdot\mathbf{k}\right)
      \left(C_8M\right)^{-1},
\end{equation}
where $R(\pi/4)$ denotes the two-dimensional rotation matrix corresponding to a counterclockwise rotation by an angle $\pi/4$, explicitly given by
\begin{equation}
    R\!\left(\pi/4\right)
    =
    \begin{pmatrix}
        \cos(\pi/4) & -\sin(\pi/4) \\
        \sin(\pi/4) & \cos(\pi/4)
    \end{pmatrix}.
\end{equation}

\section{Symmetry-Reduced Spectral Localizer}
\begin{figure}
    \label{App:reducedlocalizer}
    \centering
    \includegraphics[width=1\columnwidth]{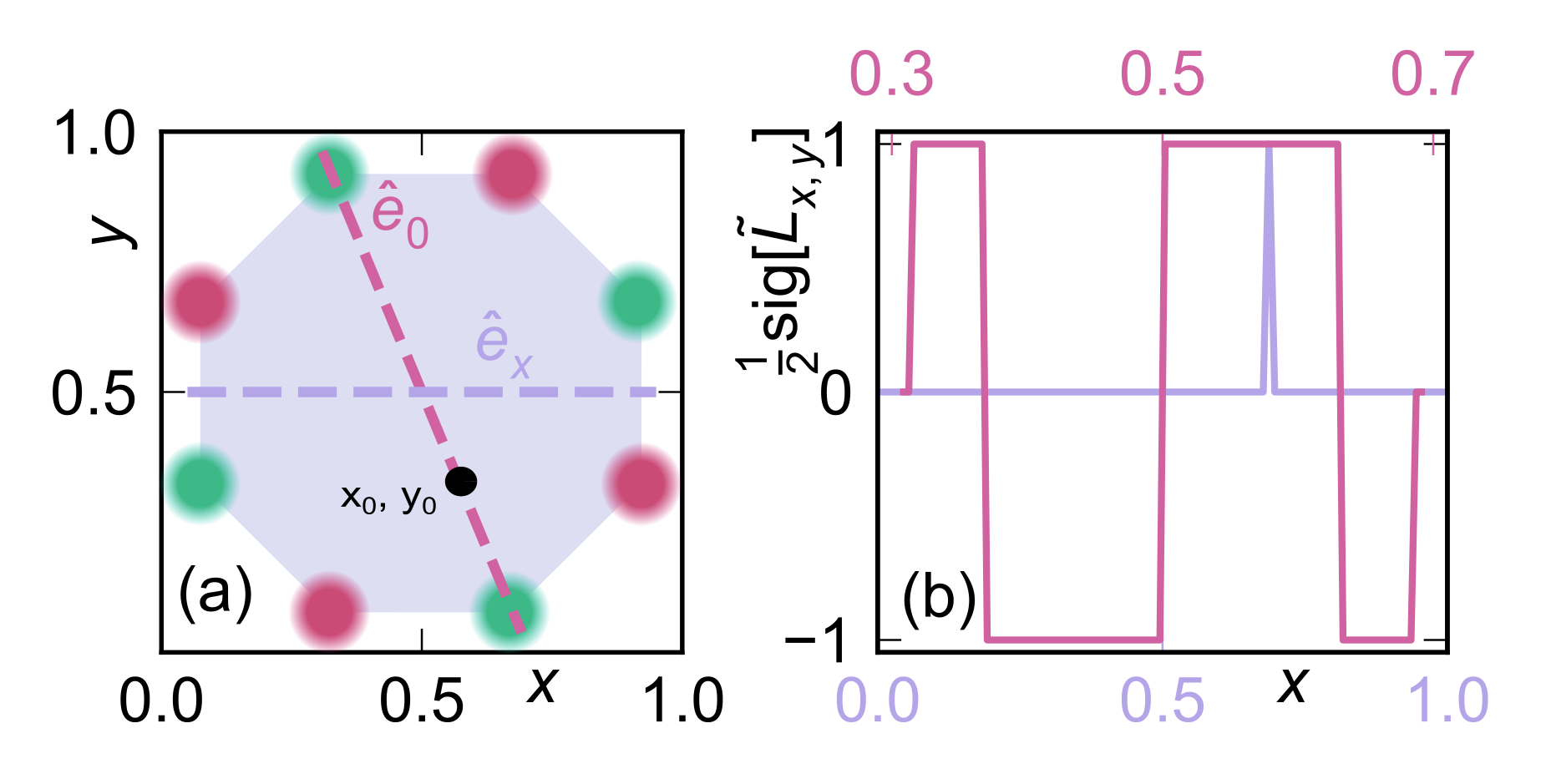}
    \caption{
        (a) Chiral charges of the zero-energy modes for $\mu/t \sim -2$, green and red circles denote positive and negative chiral charges, respectively. The vector $\hat{e}_0$ connects 
        two opposite vertices of the octagon, while $\hat{e}_x$ is the horizontal direction. We mark the point $x_0, y_0$ where we compute the invariant in \cref{fig:Fig3}(c) in the main text.
        (b) Half-signature of the symmetry-reduced localizer as a function of $x$, along direction $\hat{e}_0$ in pink and $\hat{e}_x$ in purple.
    }
    \label{fig:apx_symredloc}
\end{figure}
\begin{figure}
    \centering
    \includegraphics[width=1\columnwidth]{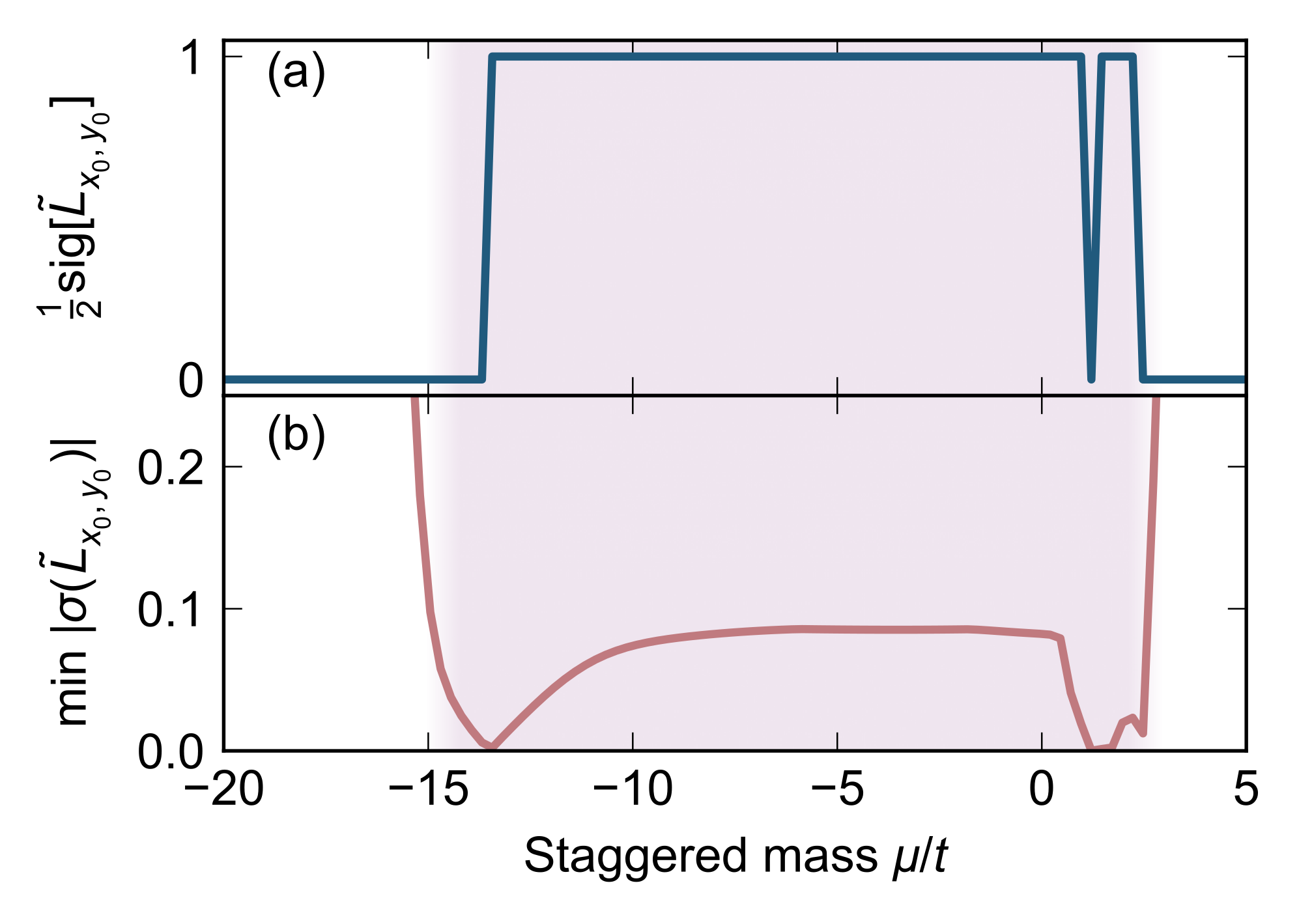}
    \caption{ (a) Half-signature of the symmetry-reduced localizer as a function of $\mu/t$ from \cref{fig:Fig3}(c), with $\kappa=0.5$, $\mathbf{x}=(x_0, y_0)$. (b) Local gap of the symmetry-reduced localizer corresponding to panel (a) as a function of $\mu/t$. The shaded region is that shown in \cref{fig:Fig3}.
    }
    \label{fig:apx_locgap}
\end{figure}
We start from the one-dimensional spectral localizer~\cite{LORING2015,Cerjan_tutorial_2024}, which can be written as
\begin{equation}
    \label{eq:apx_loc1D}
    L(E,x) = s_y \otimes (\mathcal{H}-E)  + \kappa \,s_x \otimes(X-x\1) ,
\end{equation}
where $s_i$ are Pauli matrices that act on an auxiliary subspace and $\1$ is  shorthand notation for $\1_N \otimes\sigma_0\otimes \tau_0$. For a system with even dimensionality, the spectrum of this operator, e.g. its half-signature, provides information about the topology of the system~\cite{LORING2015,Cerjan_tutorial_2024}. In this case, $L$ is off-diagonal and hermitian, which implies that its spectrum is symmetric around zero energy, thus leading to a trivial signature. To overcome the implicit redundancy when defining an invariant, one can exploit the chiral symmetry of the Hamiltonian \cref{eq:hamgen}. Let $\mathcal{S} =  \1_N \otimes\sigma_x\otimes\tau_0$ be the chiral operator, which satisfies
\begin{align}
    \mathcal S^\dagger &= \mathcal S,~~~~
    \mathcal S^2= \mathbf 1,~~~~
    \{\mathcal S,\mathcal{H}\} = 0,~~~~
    [\mathcal S, X]= 0.
    \label{eq:apx_propertieS}
\end{align}
If we apply a unitary transformation 
$ \mathcal{U}=\begin{pmatrix}
    \mathcal{S} & 0 \\
    0 & \1
\end{pmatrix}$ 
to the localizer in \cref{eq:apx_loc1D}, the relations in \cref{eq:apx_propertieS} allow us to obtain a transformed localizer of the form 
\begin{widetext}
\begin{equation}
    \label{eq:apx_unitlocalizer}
    L_U(E,x)= \mathcal{U}^\dagger L(E,x)\mathcal{U} 
    =\begin{pmatrix}
    0 & \mathcal{S}[-i(\mathcal{H} - E) + \kappa(X - x \1) ] \\
    [i(\mathcal{H} - E) + \kappa(X - x \1) ]\mathcal{S}  & 0
    \end{pmatrix}.
\end{equation}
\end{widetext}
The advantage of this form is that its spectrum coincides with that of $L(E,x)$ while each of its off-diagonal blocks contains half of the spectrum, provided the blocks are Hermitian. This condition only holds for $E=0$ as can be verified from the relations listed in \cref{eq:apx_propertieS}. Therefore, it suffices to consider only one block of $L_U(E,x)$ to extract the topological information of the system at $E=0$, which defines the \emph{symmetry-reduced localizer}
\begin{align}
    \label{eq:apx_symredloc}
    \tilde{L}_{x} = [i\mathcal{H} + \kappa(X - x\1) ]\mathcal{S}.
\end{align}
For a two-dimensional system as the one considered in this work, $X$ can be replaced by any linear combination of the position operators $X$ and $Y$ such as $\hat{e}_d \cdot (\mathbf{X}-\mathbf{x}\1)$, with $\hat{e}_d$ the direction along which we wish to project the system. 

As discussed in the main text and Ref.~\cite{Cerjan_local_2024}, changes in the half-signature of this operator signal variations in the topological phase of the system. In \cref{fig:apx_symredloc}, we show the half-signature of the symmetry-reduced localizer as a function of $x$ for $\mu/t \sim -2$, for which the system hosts eight zero-energy modes. We consider two different directions to evaluate the position operator: one choosing $\hat{e}_d$ along $\hat{e}_0$, connecting two opposite vertices of the octagon, and one choosing $\hat{e}_d$ along the $x$-axis, $\hat{e}_x$, see \cref{fig:apx_symredloc}(a).

\cref{fig:apx_symredloc}(a) displays the chiral charges of the corner modes, colored with respect to the eigenvalues of the operator $\mathcal{S}$ projected onto the zero-energy subspace. Variations in the half-signature occur precisely when the reference point $x$ crosses a zero-energy mode, with magnitude determined by its chirality $\pm 1$. Hamiltonian eigenstates away from zero-energy do not contribute to the half-signature, as they come in pairs with opposite chiral charges. Consequently, $\hat{e}_x$, as well as any direction that projects pairs of modes with reversed chirality, fails to resolve the topology, whereas $\hat{e}_0$ successfully captures the changes in the half-signature~\cite{Cerjan_local_2024,Cerjan_tutorial_2024}. This argument underlies our choice of $\hat{e}_0$ for computing the phase diagram in \cref{fig:Fig3}(c). The non-zero values observed along $\hat{e}_x$ in \cref{fig:apx_symredloc}(b) are due to small deviations from exact $C_8$ symmetry in the chiral charge distribution.

To assess the stability of the topological phase, it is helpful to analyze not only the half-signature of the symmetry-reduced localizer, but also its local gap, defined as the smallest absolute value of its spectrum~\cite{Cerjan_tutorial_2024}, $\mathrm{min}|\sigma(\tilde{L}_\mathbf{x})|$. The size of the local gap sets a lower bound for the perturbation required to close the gap; only those perturbations whose magnitude is at least as large as the local gap can drive the localizer to change its signature, thereby altering the topological phase~\cite{Cerjan_tutorial_2024}. Hence, a large non-vanishing local gap ensures that the half-signature is stable against perturbations, thus providing a robust phase.

In \cref{fig:apx_locgap}(a), we reproduce the localizer index, already shown in 
\cref{fig:Fig3}(c), and its corresponding local gap in panel (b). We observe that the gap remains open across the topological region, thereby confirming the stability of the phase. This plot also informs us that the numerical instability observed at $\mu/t \sim 0$ arises from the small real-space Hamiltonian gap at this region (see \cref{fig:Fig1}(c)), which is inherited by the localizer. In addition, the fact that the local gap remains finite also confirms that the chosen value of $\kappa$ is suitable for capturing the phase transitions.

\end{document}